\documentclass[superscriptaddress,prl,twocolumn,nofootinbib,a4paper]{revtex4}

\usepackage{amstext,amsmath,amssymb}
\usepackage[dvips]{graphicx}
\usepackage{latexsym}
\usepackage{floatflt}

\newcommand{\Ref}[1]{(\ref{#1})}

\newcommand{\Id}{\textrm{Id}}

\def\be{\begin{equation}}
\def\ee{\end{equation}}
\def\bes{\begin{eqnarray}}
\def\ees{\end{eqnarray}}

\def\arr{\rightarrow}

\def\om{\omega}

\def\w{\wedge}
\def\la{\langle}
\def\ra{\rangle}
\def\f{\frac}
\def\tl{\widetilde}

\def\hh{{\cal H}}

\def\ii{{\cal I}}

\def\cc{{\cal C}}

\def\ss{{\cal S}}
\def\aa{{\cal A}}

\def\eps{\epsilon}

\newcommand{\lalg}[1]{\mathfrak{#1}}
\newcommand{\SU}{\mathrm{SU}}
\newcommand{\SO}{\mathrm{SO}}
\newcommand{\Spin}{\mathrm{Spin}}

\newcommand{\su}{\lalg{su}}

\newcommand{\so}{\lalg{so}}

\def\eps{\epsilon}

\newcommand{\rec}{\begin{picture}(0,0)(-85,25) \put(-14,17){\Large{$=$}}
          \put( 0,0){\small{$j^+_1$}}\put( 0,35){\small{$j^+_2$}}
          \put(45,0){\small{$j^+_4$}}\put(45,35){\small{$j^+_3$}}
          \put(10,10){\line(1,1){10}}\put(10,30){\line(1,-1){10}}
          \put(35,20){\line(1,1){10}}\put(35,20){\line(1,-1){10}}
          \put(20,20){\line(1,0){15}}\put(22,25){\small{$j_{12}^+$}}
          \put(20,20){\circle*{3}}\put(35,20){\circle*{3}}
\end{picture}
\begin{picture}(0,0)(-150,25) \put(-12,17){$\bigotimes$}
          \put( 0,0){\small{$j^-_1$}}\put( 0,35){\small{$j^-_2$}}
          \put(45,0){\small{$j^-_4$}}\put(45,35){\small{$j^-_3$}}
          \put(10,10){\line(1,1){10}}\put(10,30){\line(1,-1){10}}
          \put(35,20){\line(1,1){10}}\put(35,20){\line(1,-1){10}}
          \put(20,20){\line(1,0){15}}\put(22,25){\small{$j_{12}^-$}}
          \put(20,20){\circle*{3}}\put(35,20){\circle*{3}}
\end{picture}
\begin{picture}(0,0)(0,25)
\thicklines
          \put( -10,0){\small{$(j^+_1,j^-_1)$}}\put( -10,35){\small{$(j^+_2,j^-_2)$}}
          \put(30,0){\small{$(j^+_4,j^-_4)$}}\put(30,35){\small{$(j^+_3,j^-_3)$}}
          \put(10,10){\line(1,1){10}}\put(10,30){\line(1,-1){10}}
          \put(35,20){\line(1,1){10}}\put(35,20){\line(1,-1){10}}
          \put(20,20){\line(1,0){15}}%\put(22,25){$i$}
          \put(20,20){\circle*{3}}\put(35,20){\circle*{3}}
\end{picture}  }

\begin{document}

\title{Consistently Solving the Simplicity Constraints for Spinfoam Quantum Gravity}

\author{{\bf Etera R. Livine}}\email{etera.livine@ens-lyon.fr}
\affiliation{Laboratoire de Physique, ENS Lyon, CNRS UMR 5672, 46 All\'ee d'Italie, 69007 Lyon}
%\author{{\bf Carlo Rovelli}}\email{rovelli@cpt.univ-mrs.fr}
%\affiliation{Centre de Physique Th\'eorique, CNRS UMR 6207, Campus de Luminy Case 907, 13009 Marseille}
\author{{\bf Simone Speziale}}\email{sspeziale@perimeterinstitute.ca}
\affiliation{Perimeter Institute, 31 Caroline St North, Waterloo, ON N2L 2Y5, Canada}

\begin{abstract}

\noindent We give an independent derivation of the Engle-Pereira-Rovelli spinfoam model
for quantum gravity which recently appeared in [arXiv:0705.2388]. Using the coherent state
techniques introduced earlier in [arXiv:0705.0674], we show that the EPR model realizes a
consistent imposition of the simplicity constraints implementing general relativity from a
topological BF theory.

\end{abstract}

\maketitle

%%%%%%%%%%%%%%%%%%%%%%%%%%%%%%%%%%%%%%%%%%%

In the recent years, spinfoam models have developed as a promising approach to quantum gravity
providing us with a regularized path integral formalism. They are formulated as state sums which
define transition amplitudes for almost-topological quantum field theories. Now, general relativity
(GR) in its first order formalism can be recasted as a constrained BF theory with the following
action:
\be\label{1}
S_{\rm GR}[B,\om,\lambda]\,=\int_{M} B^{IJ}\wedge F_{IJ}[\om] +\lambda_\alpha \, \cc_\alpha[B].
\ee
Notation is as follows. $M$ is the space-time manifold, $I,J$ are Lorentz indices running from 0 to
3, $\om$ is a $\so(3,1)$-valued 1-form and $F$ is its strength tensor, $B$ is a $\so(3,1)$-valued
2-form, and $\alpha$ a set of labels. Sums over repeated indices are implicit. The first term $\int
BF$ taken on its own defines a topological field theory with no-local degrees of freedom and no
geometrical interpretation. It admits a straightforward exact spinfoam quantization. The second
term consists of quadratic constraints $\cc_\alpha[B]$ enforced by the Lagrange multipliers
$\lambda_\alpha$. It reduces the number of independent components of the $B$-field so to express it
in term of 1-forms $e_I$ as (with a caveat discussed below in the last section)
$B^{IJ}=\epsilon^{IJKL}(e_K\wedge e_L)$. Through these constraints, the theory is shown to be
equivalent to GR with $e$ and $\om$ interpreted as the tetrad field and the Lorentz connection.

These $\cc[B]$ constraints are called the simplicity constraints for the bivector field $B$ and the
key issue of the spinfoam program is to implement them consistently at the quantum level in the
regularized path integral. The most studied spinfoam model up to now is the Barrett-Crane model
\cite{bc1,bc2}. It is actually the only model which has been developed enough to allow for practical calculations
and numerical simulations. Despite these advances, this model has been greatly criticized from many
perspectives and it is widely believed that it has to be substantially modified to yield a proper
spinfoam theory for quantum gravity, e.g.
\cite{laurent2,baez1,daniele1,graviton1,ls1,epr1,sergei1,matrice1}.
The most recent and most convincing criticism is that the Barrett-Crane formula does not lead to
the right spin-2 tensorial structure for the graviton propagator in the semi-classical limit
\cite{graviton1,matrice1}. This issue is traced back to the Hilbert space of boundary states being too
poor to allow such a tensorial structure for the correlation functions \cite{graviton1,ls1}.

There have been a few proposals for new models attempting to cure this problem
\cite{daniele1,ls1,epr1,sergei1}. In the present paper, we realize the program proposed
in \cite{ls1} and show that it leads to the same model recently derived by different means in
\cite{epr1,epr2}. The issue with the Barrett-Crane model is the following.
The constraints do not commute with each other, i.e. $[\hat{\cc}_\alpha,\hat{\cc}_\beta]\ne0$,
reflecting their correspondence to second class constraints in a canonical analysis \cite{sergei1,sergei2,e1}.
Nevertheless, the Barrett-Crane model implements the constraints %$\cc[B]=0$
strongly at the quantum level and identifies boundary states as satisfying
$\hat{\cc}_\alpha\,|\psi\ra=0$ for all labels $\alpha$. This is not the procedure that one ordinary
follows for non-commuting constraints, and it leads to an over-constrained Hilbert space, with not
enough degrees of freedom to describe a 3-geometry. It is more natural to impose them in a weaker
way, using for instance coherent states.
This is similar to identifying single particle quantum states satisfying $x=p=0$: there is no state
exactly solving $\hat{x}\,|\psi\ra=\hat{p}\,|\psi\ra=0$ and we instead use coherent states
satisfying these conditions in expectation values with minimal uncertainty.
%[\emph{maybe not super enlightening! shall we cut it? or spend an extra line for this analogy?}]
This leads to a larger Hilbert space which will hopefully have a better-behaved semiclassical
sector.

We will first review the structure of the discretized $\cc[B]$ constraints in the spinfoam
framework and show how they truly implement the second class constraints derived in the canonical
analysis. We then show how to impose them weakly at the quantum level and derive the new Hilbert
space of boundary states. We finally explain how to implement this idea using the coherent intertwiner
states introduced in \cite{ls1}. This leads to a new proposal for a spinfoam model of quantum
gravity.

We point out that we work in 4d Riemannian gravity with spacetime signature $(++++)$ and gauge
group $\SO(4)$. Even though some of the key ideas will extend directly to Lorentzian signature and
the non-compact group $\SO(3,1)$, we postpone a detailed discussion of this case for later
work.

%We do not think that moving to the Lorentzian case will affect the logic we follow but dealing with
%the non-compact Lorentz group instead of the compact group $\SO(4)$ most likely requires more a
%careful mathematical analysis.

%Finally, there might be considerable overlap with \cite{epr2,emmerdeurs} whose authors have been
%parallelly working on the same problematic. {\bf should we include this sentence?}

%%%%%%%%%%
\section{The Simplicity Constraints}
%\section{Simplicity Constraints for the 4-Simplex}
%%%%%%%%%%
Our starting point is a discretization of \Ref{1} on a simplicial manifold representing spacetime.
This is made of 4-simplices glued along common tetrahedra. Each 4-simplex has five tetrahedra and
ten triangles. The fields $B$ and $\om$ are then discretized and quantized, e.g.
\cite{alej,michael}. We will focus on the $B$ field since our purpose is to show how to deal with
the constraints $\cc[B]$. At the quantum level, a representation of $\SO(4)$ is associated to each
triangle $\Delta$ and the variables $B^{IJ}_\Delta$ are represented as the $\so(4)$-generator
$J^{IJ}$ in that representation. Then gauge invariance allows us to associate a quantum state to
each tetrahedron, given by the intertwiner between the four representations attached to its four
boundary triangles, i.e a $\SO(4)$-invariant state in the tensor product of these four
representations. Notice that in this procedure a tetrahedron state is uniquely defined by the
tetrahedron irrespective to the 4-simplex to which it belongs. Finally, a quantum 4-simplex
consists in the ten representations labeling its triangles and the five quantum states associated
to its tetrahedra. Tensoring these tetrahedron states and tracing out over the representations, we
get a scalar amplitude for each 4-simplex. The spinfoam amplitude is defined as the product of
these 4-simplex amplitudes.

The topological BF theory is obtained by allowing all irreducible (unitary) representations for triangles
and all intertwiner states for tetrahedra. A constrained BF theory such as gravity restrains both the representations
and the intertwiner spaces. For instance, the Barrett-Crane model uses the simple representations
of $\Spin(4)$ and the unique Barrett-Crane intertwiner. So there are no degrees of freedom
in the intertwiner space. Here we will relax the way of imposing the
simplicity constraints in order to enlarge the intertwiner space.

The $\cc[B]$ constraints usually read for all space-time indices (greek letters):
\be\label{2}
%\eps B_{\mu\nu} B_{\rho\sigma} \equiv
\eps_{IJKL}B^{IJ}_{\mu\nu}B^{KL}_{\rho\sigma}
=\eps_{\mu\nu\rho\sigma}\,\f b{4!},
\ee
with $b=\epsilon^{IJKL}\epsilon_{\mu\nu\rho\sigma}B_{IJ}^{\mu\nu}B_{KL}^{\rho\sigma}$. They ensure
that $B$ comes from a tetrad field $e$ \cite{laurent3}. At the canonical level, \Ref{2} translates
into second class constraints: a set of primary constraints ensuring that the relation between $B$
and $e$ holds on the canonical hypersurface, plus a set of secondary constraints ensuring that it
also holds under time evolution. Both sets of constraints are essential to compute the Dirac
bracket on the phase space \cite{sergei3}. A criticism of the spinfoam quantization is that it
seems to take into account only the primary constraints \cite{sergei1,sergei2,e1,sergei3}. We
address this issue below, and identify the secondary constraints. Notice also that in the
Lorentzian case the secondary constraints correspond to the reality constraints of self-dual loop
gravity, so it would be enlightening to understand how spinfoams deal with them.
\begin{floatingfigure}[r]{3cm}\includegraphics[width=2cm]{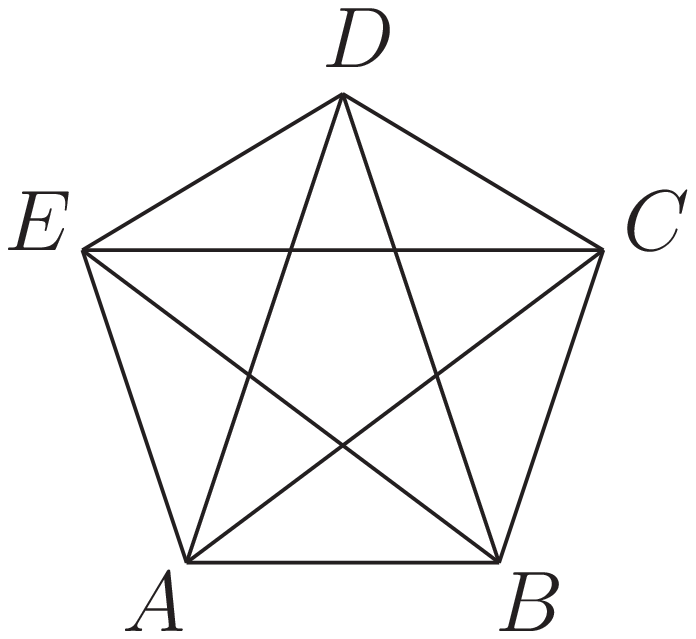}  %\caption{The dual 4-simplex.}
\end{floatingfigure}
Since the building elements of spinfoams are the 4-simplices, we now look in details at the
simplicity constraints within each 4-simplex. We call $A,B,C,D,E$ the five tetrahedra of the
4-simplex. The adjacient picture shows the dual 4-simplex.
Each of the ten triangles is labeled by a pair of tetrahedra sharing it, e.g. $(AB)$.
Consider the discrete variables $B^{IJ}_\Delta\equiv B^{IJ}_{AB}$ within the 4-simplex.
First, they are constrained to satisfy a
closure condition for each tetrahedron, namely $B_{AB}+B_{AC}+B_{AD}+B_{AE}=0$
for the tetrahedron $A$, and so on. This is the
discrete equivalent of the Gauss law ensuring the $\SO(4)$ gauge invariance. Then the constraints
$\cc[B]$ impose further conditions on these $B$ variables. These equations are labeled by couples
of triangles $(\Delta,\Delta')$ and we distinguish three different cases \cite{laurent3,michael}:
\begin{itemize}
\item $\cc^{(1)}$: when $\Delta=\Delta'$, the associated bivector $B_\Delta$ must be simple, $\eps_{IJKL}B^{IJ}_\Delta
B^{KL}_{\Delta}=0$.
\item $\cc^{(2)}$: when $\Delta$ and $\Delta'$ belong to the same tetrahedron, i.e. when they share
a common edge, we also have $\eps B_\Delta B_{\Delta'}=0$. This means that the sum $B_\Delta +B_{\Delta'}$ of the two
bivectors is once again simple.
\item $\cc^{(3)}$: when $\Delta$ and $\Delta'$ only share a common vertex, they don't belong to the same
tetrahedron. The constraints impose that the quantity $\eps B_\Delta B_{\Delta'}$ is, up to a sign,
independent of the choice of such couple of triangles. It is actually equal to the 4-volume of the
4-simplex (up to a factor 16/25) and the sign relates to the relative orientation of the triangles.
\end{itemize}

The case $\cc^{(1)}$ is straightforward to deal with. The two remaining cases are the problematic
ones. We naturally would like to interpret $\cc^{(2)}$ as the primary constraints and $\cc^{(3)}$
as the secondary constraints. Assuming that $\cc^{(2)}$ holds on the initial hypersurface (e.g. one
tetrahedron of the 4-simplex) and that $\cc^{(3)}$ are satisfied, then $\cc^{(2)}$ is also true on
the final hypersurface (e.g. all four remaining tetrahedra). This is easily proved using the
closure condition to relate $\cc^{(2)}$ and $\cc^{(3)}$. For instance,
\be
\eps B_{AB}B_{BC} +\eps B_{AC} B_{BC} = -\eps B_{AD}B_{BC} -\eps B_{AE}B_{BC},
\ee
where the subscripts $I,J,K,L$ are implicit. The left hand side corresponds to $\cc^{(2)}$ on the
tetrahedra $B$ and $C$ while the right hand side relates to $\cc^{(3)}$. We then repeat this
procedure on all tetrahedra. This answers the question raised above: $\cc^{(3)}$ are the secondary
constraints searched for: they involve the whole 4-simplex structure and ensure that the spatial
constraints $\cc^{(2)}$ are satisfied under time evolution.

Furthermore, using the same relations, we show that assuming $\cc^{(2)}$ holds for all tetrahedra
of the 4-simplex implies that $\cc^{(3)}$ is true. This means that we only need to solve the
constraints $\cc^{(2)}$ as suggested in \cite{e1}. This is also the reason why the case
$\cc^{(3)}$ is not discussed in the geometric characterization of 4-simplices in the original
Barrett-Crane papers \cite{bc1,bc2}.

At the quantum level, we replace all the variables $B^{IJ}_\Delta$ by the $\so(4)$ generators
$J^{IJ}_\Delta$. This does not change anything to the previous statements. The issue is that,
although the constraints $\cc^{(1)}$ commute with each other, the constraints $\cc^{(2)}$ (and
$\cc^{(3)}$) do not. Therefore looking for states that solve exactly all the constraints
$\cc^{(2)}$ might lead to a very small Hilbert space. Indeed it gives the unique Barrett-Crane
intertwiner. However this situation comes from the fact that these constraints are second class
already at the classical level.
%, i.e their Poisson brackets do not vanish.
This suggests a different approach: to weaken the constraints and look for coherent states that
would only solve them in average with a minimal uncertainty \cite{ls1}. This should likely lead to
states with a more straightforward geometrical interpretation and provide us with a larger Hilbert
space.

%%%%%%%%%%
\section{Enlarging the Hilbert Space}
%%%%%%%%%%
In this section we construct a larger Hilbert space, where the simplicity constraints hold
in the expectation values.
Consider $\cc^{(1)}$ first. In terms of generators, the constraint on a single triangle $\eps
J_\Delta J_\Delta = 0$ is a condition on the $\so(4)$ representation associated to $\Delta$. To
understand this condition, notice that $\eps J J$ is the second Casimir operator of the $\so(4)$
Lie algebra; using the decomposition of $\so(4)$ in self-dual and anti-self-dual sectors,
$\so(4)=\su_+(2)\oplus\su_-(2)$, it is the difference of the Casimirs of the two $\su(2)$
sub-algebras:
\be
\eps_{IJKL}J_\Delta^{IJ}J_\Delta^{KL}=
(\vec{J}^+_\Delta)^2 -(\vec{J}^-_\Delta)^2=0.
\ee
This means that the $\so(4)$ representation $(j^+,j^-)$ associated to the triangle $\Delta$ must carry the same spin on
its self-dual and anti-self-dual part, $j^+=j^-$. Such a representation is called
simple \cite{bc1,laurent4}.

Next, consider $\cc^{(2)}$, and notice that it involves two triangles on the same tetrahedron. We have four simple
representations $(j_a,j_a)$ for the four triangles $\Delta_{a=1..4}$ on
the tetrahedron boundary. The closure condition $\sum_a J_a=0$ means that we are restricted to
$\so(4)$-invariant states in the tensor product $\hh\equiv\,\otimes_a \hh_{(j_a,j_a)}$, i.e.
intertwiner states between these four representations.
We use the standard recoupling basis of intertwiners,

\vspace{0.7cm} \rec \vspace{1.2cm}

\noindent and we label
$|j_a, (j_{12}^+,j_{12}^-)\ra$ the states in $\hh$, where $j_a$ denotes the simple representation $(j_a,j_a)$, and
$(j_{12}^+,j_{12}^-)$ is the label for the representation $(J_1+J_2)$.
In $\hh$ we have three new
independent simplicity conditions, $\cc_{a,b}\,\equiv\,\eps J_a J_b=0$ for all couples of triangles
$(\Delta_a,\Delta_b)$. These constraints mean that the sum $(J_a+J_b)$ is required to remain
simple.
Strongly imposing the simplicity conditions $\cc_{1,2}$ forces the recoupled
representation to be simple, $j_{12}^+=j_{12}^-$. Further imposing $\cc_{1,3}$ and $\cc_{1,4}$ then
leads to a single intertwiner \cite{michael2}.
The key point is that these constraints do not commute with each other. For instance
$[\cc_{1,2},\cc_{1,3}]$ is still cubic in the $J$'s \cite{michael2,bb}. Thus, imposing these
constraints strongly at the quantum level amounts to imposing a whole tower of constraints of higher and
higher order in the $J$'s. It looks as if we are actually enforcing too many conditions, and we are
indeed left with a one-dimensional intertwiner space (once the $j_i$'s are given).

We propose to weaken the constraints and look for intertwiner states $\psi$ that satisfy the
simplicity conditions only in the expectation values, $\la \psi|\cc_{a,b}|\psi\ra\,=0$ for all couples
$(a,b)$. For this purpose, we introduce the Hilbert space of {\it symmetric} intertwiners,
$\hh^{\rm sym}_0$. These are defined as invariant under the exchange of $j_{12}^+$ and $j_{12}^-$:
\be\label{5}
|\psi\ra\,=\sum_{(j_{12}^+,j_{12}^-)}\,
\psi_{j_{12}^+,j_{12}^-}
\,|j_a, (j_{12}^+,j_{12}^-)\ra,
\ee
with $\psi_{j^+,j^-}=\psi_{j^-,j^+}$. It is straightforward to check that this defines a Hilbert
space and that it is invariant under the choice of recoupling basis -- in \Ref{5} we could have
chosen the pairing 1-3 or 1-4 instead of 1-2. It is also obvious that any state $\psi\in\hh^{\rm
sym}_0$ satisfies all the constraints $\cc_{a,b}$ in expectation value. We even have the stronger
statement that $\hh^{\rm sym}_0$ is the largest Hilbert space such that all the matrix elements of
the constraints vanish:
\be
\forall \phi,\psi\in\hh^{\rm sym}_0,\quad
\la \phi|\cc_{a,b}|\psi\ra\,=0.
\ee
From this perspective, the operators $\cc_{a,b}$ can take us out of the Hilbert space
$\hh^{\rm sym}_0$, but they actually vanish weakly if we restrict ourself to work only with states within
$\hh^{\rm sym}_0$.
Although $j_{12}^+$ and $j_{12}^-$ are not necessarily equal, the simplicity condition is guaranteed by
the symmetry of the coefficients.

At the end of the day, we have shown that it is possible to consistently impose the intertwiner
simplicity condition in a weaker sense. This leads a larger intertwiner space, thus a larger space
of (boundary) spin networks for the spinfoam model. On this larger space the simplicity condition holds in
average. In the next section, we show how to impose the constraints with (almost) minimal uncertainty using the
coherent intertwiners introduced in \cite{e1}. This allows to recover the geometrical interpretation
of intertwiners as quantum tetrahedra.

%%%%%%%%%%
\section{Coherent Simple Intertwiners}
%%%%%%%%%%

Let us start by considering a bivector $B_\Delta$ associated to a single triangle. It is simple if
and only if its self-dual and anti-self-dual parts have equal norms, $|\vec{b}^+|=|\vec{b}^-|$,
with $b^\pm_i\,\equiv\,(B_i\pm B^{0i})/2$ and $B_i\,\equiv\,\f12\eps_{ijk}B^{jk}$ is the spatial
part of the bivector.

The simpler case when $\vec{b}^+=\vec{b}^-$ means that the ``time-like" part of $B$ vanishes,
$B^{0i}=0$, i.e the ``time-like" vector $N^{(0)}=(1,0,0,0)$ is orthogonal to $B$.\footnote{The
notion of ``time-like" is not properly defined in the Euclidean space, and furthermore there are
actually two 4-vectors orthogonal to any given simple bivector. The vector $N^{(0)}$ can
nevertheless naturally be seen as the ``time-like" normal vector. This issue would be clearer in a
Lorentzian framework which we postponed for future investigation.} Then the spatial part $\vec{B}$
can always be expressed as the vector product of two 3-vectors $\vec{e}$ and $\vec{f}$,
$B_i=\eps_{ijk}e_j f_k$. Defining the 4-vectors $e=(0,\vec{e}\,)$ and $f=(0,\vec{f}\,)$, it is
straightforward to check that $B^{IJ}=e^{[I}f^{J]}$, that is the bivector can be expressed as the
wedge product of two vectors which are interpreted as a discretized tetrad field.

In the generic case, if $\vec{b}^+$ and $\vec{b}^-$ have the same norm, there exists a $\SO(3)$
rotation $g$ which maps one on the other, $\vec{b}^-\,=\,g\,\vec{b}^+$. Introducing the $\SO(4)$
rotation $G=(g,\Id)$ (where the left side acts as $\SU(2)_+$ and the right on $\SU(2)_-$), we
define the rotated bivector $\tl{B}\,\equiv\,G^{-1}BG$. Then $\tl{B}$ has equal self-dual and
anti-self-dual components and we can repeat the same analysis as above. In particular, we obtain
that the 4-vector $N\,\equiv\,GN^{(0)}$ is the ``time-like" vector orthogonal to $B$.

We now discuss the implementation of this idea at the quantum level.
Following \cite{e1}, we introduce a coherent state which is peaked on the classical value $B^{IJ}$.
Such state is the tensor product of two $\SU(2)$ coherent
states for the self-dual and anti-self-dual components, $|j^+, \hat{n}^+, j^-,\hat{n}^-\ra$, where
$\vec{b}^\pm=j^\pm\hat{n}^\pm$ and the $\hat{n}^\pm\in\ss^2$ are unit 3-vector.
Satisfying the simplicity condition $\cc^{(1)}$ means
choosing the same representation for both components, $j^+=j^-$,
which we denote simply as $j$. The $\SO(3)$ rotation $g$ between $\hat{n}^+$ and $\hat{n}^-$
defines the time-like normal to $B$ as discussed above.

A tetrahedron is characterized by four bivectors $B_{a}$, $a=1\ldots 4$, each of which has
associated a coherent state $|j_a,\hat{n}^+_a,\hat{n}^-_a\ra$, satisfying the closure condition
$\sum_a B_a=0$. A quantum state for the tetrahedron is then constructed by averaging over
$\Spin(4)$ the tensor product of the four coherent states for each bivector:
$$
\int_{\Spin(4)} dG\, \otimes_{a=1}^4\,G\,|j_a,\hat{n}^+_a,\hat{n}^-_a\ra.
$$
The $\Spin(4)$-averaging ensures the state is an intertwiner, so as to satisfy the closure
constraint at the quantum level. We still have to solve the simplicity constraint. Following an
idea of \cite{epr1,sergei1}, we implement them by requiring that all four bivectors $B$ lay in the
same hypersurface: they must be normal to the same ``time-like" vector. This means that all four
self-dual components $\hat{n}^-_a$ must come from the same rotation of the four anti-self-dual
components $\hat{n}^+_a$. Thus, there must exist a single rotation $g\in\SO(3)$ independent from
$a$ such that:
\be
\forall a,\quad \hat{n}^-_a\,=\,g\,\hat{n}^+_a.
\ee
Implementing this condition on the intertwiner state, we are left with the quantum tetrahedron
state:
\be
\psi\,=\,\int dG\, \otimes_a\,G\,|j_a,\hat{n}^+_a,g\hat{n}^+_a\ra.
\ee
Since the Haar measure on $\Spin(4)$ is the product of the independent integrations over $\SU(2)_+$
and $\SU(2)_-$, the rotation $g$ is irrelevant and $\psi$ is simply a tensor
product state:
\be
\psi=\ii_+\otimes\ii_+,\quad
\ii_+\,\equiv\int_{\SU(2)}dg_+ \otimes_a \,g_+|j_a,\hat{n}^+_a\ra,
\ee
where $\ii_+$ is an $\SU(2)$-intertwiner state. Expressed as such, it is manifest that the state
$\psi$ belongs to the Hilbert space $\hh^{\rm sym}_0$ constructed above, and therefore
solves the simplicity constraints weakly.

Using the tensoring properties of the $\SU(2)$ coherent states, we have $|j,\hat{n}\ra^{\otimes
2}=|2j,\hat{n}\ra$, and thus we can simplify the formula above by doubling the spins $j_a$:
\be\label{psi}
\psi\,=\,\int dG\, \otimes_a\,G\,|2j_a,\hat{n}^+_a\ra.
\ee
This shows that our states are the same ones as defined by Engle, Pereira and Rovelli for their new
spinfoam model \cite{epr1,epr2}. The states \Ref{psi} span a Hilbert space of intertwiners which
(i) is a subspace of $\hh^{\rm sym}_0$ and therefore weakly solves the simplicity constraints, and
(ii) matches the Engle-Pereira-Rovelli proposal. We have only expressed their intertwiner space in
a different (overcomplete) basis using coherent states.

%\medskip

Since we use the same simple representations and same intertwiner spaces, we end up with the same
spinfoam model as in \cite{epr1,epr2}, which has the same boundary Hilbert space as Loop Quantum
Gravity.
Our 4-simplex amplitude is obtained by gluing five tetrahedron states together along ten triangles:
\be\nonumber
\aa_{\sigma}\,\equiv\,
\left[\int_{\SU(2)} [dg]^{\otimes 5}\,
\prod_{\Delta=1}^{10} \la j_\Delta \hat{n}_{s(\Delta)}|g_{s(\Delta)}^{-1}g_{t(\Delta)}|j_\Delta
\hat{n}_{t(\Delta)}\ra\right]^2
\ee
where $s(\Delta)$ and $t(\Delta)$ label the two tetrahedra to which the triangle $\Delta$ belongs.
Notice that, since we use a different intertwiner basis, our 4-simplex amplitude is not expressed
in term of $\{15j\}$'s as in \cite{epr1,epr2}, although the whole spinfoam amplitude should
ultimately be the same. The difference lays in the boundary data: our coherent spin network states
carry more information and have a simpler semiclassical behavior for large spins. This is to be
compared to coherent states for the harmonic oscillator which are labeled by two real numbers
instead of a single integer but that admit a straightforward semiclassical interpretation. We
expect this choice of basis to improve the geometrical interpretation of the model and the study of
its semiclassical limit.

%%%%%%
\section{A Sign Ambiguity}
%%%%%%
In this final section, we comment on an alternative model that can be constructed, using a sign
ambiguity present in our procedure. This is related to the existence of two sectors of the
constrained BF theory, e.g. \cite{laurent3,bb,daniele2}. Indeed, $B^{IJ}=\epsilon^{IJKL}(e_K\w
e_L)$ is not the only classical solution of the simplicity constraints \Ref{2}, but also
$B^{IJ}=e^{[I}\w e^{J]}$ solves them. The first solution gives a sector that reproduces general
relativity, while the second solution leads to a non-geometrical theory (the tetrad $e$ is still
required to be compatible with the connection, $d_\om e=0$, but does not necessarily satisfy the
Einstein equations). Of course, the goal is to build a spinfoam model representing the
gravitational sector and not the non-physical one.

This ambiguity is present in our framework, where it translates into a sign ambiguity. Considering
a single bivector $B$ satisfying the simplicity condition $|\vec{b}^+|^2=|\vec{b}^-|^2$, there
exists a rotation $g\in\SO(3)$ such that $\vec{b}^-=\,g\vec{b}^+$ as we considered, but we can also
flip the sign and consider the other branch defined by $\vec{b}^-=\,-g\vec{b}^+$. The first branch
corresponds to bivectors which read $B^{IJ}=e^{[I} f^{J]}$ while the second branch gives
$B^{IJ}=\epsilon^{IJ}{}_{KL} e^K f^L$, where $e$ and $f$ are two 4-vectors. This flipping
possibility clearly corresponds to the previous ambiguity.

The point is that this sign ambiguity is due to the invariance of the quadratic simplicity
constraints under the change $B^{IJ}\arr \eps^{IJ}{}_{KL}B_{KL}$. However, our way to implement the
simplicity constraints on intertwiners is not invariant under the Hodge operator $\eps^{IJ}{}_{KL}$
and should in principle distinguish the two sectors. This should be a great improvement on previous
spinfoam models.

At the quantum level, this means considering coherent states
$|j,\hat{n}^+\ra\otimes|j,-g\hat{n}^+\ra$ instead of $|j,\hat{n}^+\ra\otimes|j,g\hat{n}^+\ra$. At
the level of a single triangle, this does not make a difference since $-g\hat{n}^+$ is as good a
unit vector as $g\hat{n}^+$ in our Riemannian setting (in the Lorentzian setting, the two branches
can be distinguished, one vector belonging to the upper time-like unit hyperboloid and the other to
the lower hyperboloid). Nevertheless, using this choice to form tetrahedron states, we
end up with a different class of intertwiners:
\bes
\psi&=&\int dG\, \otimes_a\,G\,|j_a,\hat{n}^+_a,-\hat{n}^+_a\ra \\
&=& \int dg_+\, \otimes_a g_+|j_a,\hat{n}^+_a\ra \otimes \int dg_-\, \otimes_a
g_-|j_a,-\hat{n}^+_a\ra. \nonumber
\ees
Instead of tensoring the $\SU(2)$ intertwiner with itself, we tensor it with its complex conjugate.
It then leads to a slightly different 4-simplex amplitude where the coherent intertwiners labeling the
anti-self-dual part are the dual of the self-dual part instead of being identical.

The first proposal with matching self-dual and anti-self-dual intertwiners reproduces the model
proposed by Engle-Pereira-Rovelli \cite{epr1,epr2} while this second flipped model looks more like
a coherent state version of the Barrett-Crane model (which uses vanishing spin states of the
$|j,m\ra\otimes |j,-m\ra$ type with conjugate self-dual and anti-self-dual components).  The
natural question is which of the two models correspond to the proper spinfoam quantization of
general relativity, if any.

\medskip

Now that all the foundations have been set and the simplicity constraints consistently implemented,
the next step is to study the asymptotics of the new proposed spinfoam vertex and check that the
graviton propagator (e.g. \cite{graviton1}) is better behaved than for the Barrett-Crane model.
Notice that a calculation of the graviton tensorial structure will allow to discriminate between
the two proposed models, with identical or conjugate intertwiners, and check which one has the
right degrees of freedom.

%%%%%%%%%%%%%%%%%%%%%%%%%%%%%
\section*{Acknowledgements}

We would like to thank Carlo Rovelli, Roberto Pereira and Jonathan Engle for many useful
discussions, especially on the sign ambiguity. We are also grateful to Laurent Freidel for his
encouragements to finish writing up this paper. After completing the present work, we realized that
similar ideas had been developed independently by
%Freidel and Krasnov
in \cite{laurent5}.

%%%%%%%%%%%%%%%%%

\end{document}